\documentclass{elsart}

\usepackage{graphicx}
\begin{document}

\begin{frontmatter}

% Title, authors and addresses

% use the thanksref command within \title, \author or \address for footnotes;
% use the corauthref command within \author for corresponding author footnotes;
% use the ead command for the email address,
% and the form \ead[url] for the home page:
% \title{Title\thanksref{label1}}
% \thanks[label1]{}
% \author{Name\corauthref{cor1}\thanksref{label2}}
% \ead{email address}
% \ead[url]{home page}
% \thanks[label2]{}
% \corauth[cor1]{}
% \address{Address\thanksref{label3}}
% \thanks[label3]{}

\title{Signature of Quantum Chaos in SQUID's}

\author{Takeo Kato\corauthref{cor1}},
\ead{kato@a-phys.eng.osaka-cu.ac.jp}
\author{Ken-ichi Tanimoto},
\author{Katsuhiro Nakamura}
\corauth[cor1]{Corresponding author.}
\address{Department of Applied Physics, 
Osaka City University, Sumiyoshi-ku,
Osaka 558-8585, Japan}

% use optional labels to link authors explicitly to addresses:
% \author[label1,label2]{}
% \address[label1]{}
% \address[label2]{}
%\author{}
%\address{}

\begin{abstract}
Spectral statistics in band structures is studied in
a realistic model describing superconducting quantum
interference devices (SQUID's).
By controlling an external magnetic flux, the level statistics may show
a crossover from the GUE to the GOE. Effects of secondary discrete
symmetries seen in specific regions of the first Brillouin zone 
are also discussed.
\end{abstract}

\end{frontmatter}

% main text
Recent development in physics of quantum chaos has its origin
in studies of spectral statistics of quantum systems whose corresponding
classical dynamics exhibits chaos. It is now well
known that statistical properties of energy spectra are universal
in various quantum systems including, e.g., atoms, nuclei, and 
electron billiards as well as classical wave-mechanical systems
such as microwave cavities~\cite{Stockmann99}.  
Recently, spectral statistics in band structures in extended periodic
systems has been studied for driven systems~\cite{Miyazaki94}, 
electronic systems~\cite{Mucciolo94,Silberbauer95,Xu98,Xu01}
and photonic crystals~\cite{Gumen02}. Band structures have also
been studied in the context of avoided band
crossings~\cite{Ketzmerik00}. Most of these band-structured systems,
however, are sensitive to the accompanying impurities and/or
inhomogeneity. It is highly desirable to find corresponding
systems robust against the extrinsic randomness.

In this Letter, we consider superconducting quantum 
interference devices (SQUID's), whose periodic features are
described by a few macroscopic
variables called superconducting phases. 
We here focus on small SQUID's with three junctions where a
two-dimensional periodic potential for the phases can be realized.
Because of the macroscopic nature of the phases, the potential
is robust against impurities and/or inhomogeneities
of the system. This feature is of great advantage to study
the spectral statistics; This is particularly
in contrast to the two-dimensional
electron systems where impurity scattering and inhomogeneity
are unavoidable. The SQUID system also has the advantage of 
controllability.
Potential shapes for the phases can be controlled
by an external magnetic flux, and the Bloch wave-numbers can be chosen
by voltage gates. In the limited context of application to quantum
computations, one is concerned only with the lowest few quantum
levels~\cite{Orlando99,Makhlin01,van-der-Wal00,Chiorescu03}. 
Moreover generally, however, if level spectroscopy at
a high-energy region will become possible, this system would
give a good stage for studies of quantum chaos in band structures.

Already we investigated the classical phase dynamics of SQUID's
with two junctions and discovered the mysterious normal diffusion
phenomena there: The diffusion coefficient was found
to take a fractal-like feature~\cite{Tanimoto02}.
SQUID's with two junctions, however, were on scale of $100 \mu {\rm m}$
and thereby classical. On the contrary, SQUID's with three junctions
considered in this Letter is on scale of $1 \mu {\rm m}$ and their
quantization is highly desirable to elucidate a quantum signature
of the underlying mysterious normal diffusion phenomena.

\begin{figure}[tb]
\begin{center}
\includegraphics[width=80mm]{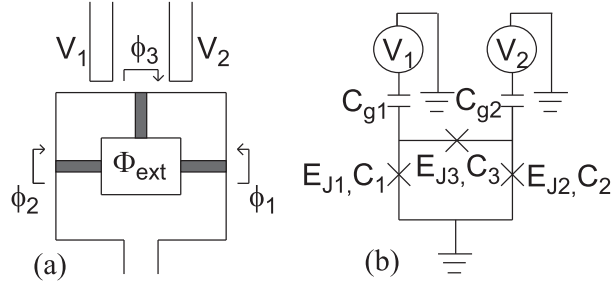}
\end{center}
\caption{(a) The three-junction system considered in this Letter.
The states of the SQUID are controlled by a magnetic flux $\Phi_{\rm ext}$
and gate voltages, $V_i$'s.
The flux is vertical to the page.
(b) The equivalent circuit of the SQUID.
The Josephson energy and capacitance of the $i$-th junction 
(shown by $\times$) are denoted by $E_{{\rm J}i}$ and $C_i$, respectively. 
The two upper superconducting 
islands are coupled to the voltage gates through
capacitances, $C_{{\rm g}1}$ and $C_{{\rm g}2}$, respectively.}
\label{fig:model}
\end{figure}

The configuration of the SQUID system is shown in
Fig.~\ref{fig:model}~(a), while the equivalent circuit is shown in 
Fig.~\ref{fig:model}~(b). The loop of superconductor includes three
Josephson junctions denoted with $\times$ in the circuit. 
The Josephson energy and capacitance of the $i$-th junction
are denoted with $E_{{\rm J}i}$ and $C_i$, respectively.   
The two upper superconducting islands are coupled to the voltage gates through
the capacitances, $C_{\rm g1}$ and $C_{\rm g2}$, respectively.
The states of this SQUID are controlled by an external flux penetrating 
through the loop, $\Phi_{\rm ext}$ and by gate voltages, $V_i$'s.

By controlling fabrication conditions,
both the Josephson energies and capacitances can be varied.
In this Letter, for simplicity we assume 
$E_{{\rm J}1}=E_{{\rm J}2}=E_{{\rm J}3} (\equiv E_{\rm J})$. 
We also assume $C_1, C_2, C_3 
\ll C_{{\rm g}1}= C_{{\rm g}2}(\equiv C_{\rm g})$, 
where the junction capacitances can be neglected. 
These assumptions are not essential to the discussion
of the classical-quantum correspondence through spectral statistics.
We assume these conditions because it is convenient to study
effects of discrete symmetries such as parity. 

Behaviors of the Josephson junction are described by the 
superconducting phase difference at the junction. We denote the phase
difference at each junction with $\phi_i$. The directions
to define these differences are shown by the arrows
in Fig.~\ref{fig:model}~(a). Then, the Hamiltonian of 
this SQUID consists of three parts:
\begin{equation}
H = H_{\rm C} + H_{\rm J} + H_{\rm L},
\label{Ham0}
\end{equation}
where $H_{\rm C}$, $H_{\rm J}$ and $H_{\rm L}$ are a 
charging energy, a junction energy and an inductance energy,
respectively. First, we show that the latter two parts 
correspond to a potential energy for the phase differences.
The junction energy is determined by the phase differences as
\begin{equation}
H_{\rm J} = - E_{\rm J} \cos \phi_1 - E_{\rm J} \cos \phi_2
- E_{\rm J} \cos \phi_3, 
\end{equation}
while the inductance energy is determined by the persistent
current along the loop, and given as
\begin{equation}
H_{\rm L} = \frac{\Phi_0^2}{2L} \left(
\phi_1-\phi_2-\phi_3-2\pi \frac{\Phi_{\rm ext}}{\Phi_0} \right)^2.
\end{equation}
Here, $\Phi_0=h/2e$ is a unit flux, and $L$ is an inductance 
determined by the size and shape of the loop. Here, we consider a small loop
with a small inductance satisfying $\Phi_0^2/2L \gg E_{\rm J}$.
Such SQUID can be easily realized experimentally by fabricating
the superconducting loop with the size of order of micrometers. 
In this small SQUID, the inductance energy approximately 
gives a constraint $\phi_1-\phi_2-\phi_3-2\pi \Phi_{\rm ext}/\Phi_0 = 0$
on the phases. By using this constraint,
the potential energy for the phases are obtained as
\begin{equation}
H_{\rm J} = - E_{\rm J} \cos \phi_1 - E_{\rm J} \cos \phi_2
- E_{\rm J} \cos (\phi_1-\phi_2-2\pi f),
\end{equation}
where $f = \Phi_{\rm ext}/\Phi_0$ is a normalized external flux.
This two-dimensional periodic potential has
a square unit cell defined by $0\le \phi_1 \le 2\pi$ and
$0 \le \phi_2 \le 2 \pi$.

On the other hand, the charging energy $H_{\rm C}$ is described by the 
charges of the junctions. We denote the charge induced at the
$i$-th gate capacitance by $Q_i = - 2e n_i$, where $n_i$ is the number
of the induced Cooper-pair. It is known that the number
of the Cooper pairs, $n_i$ is conjugate to the phase difference 
$\phi_i$ and the following exchange relation satisfies:
\begin{equation}
[ \phi_i, n_j ] = {\rm i} \delta_{ij}, \hspace{5mm}
(1\le i,j \le 2). 
\end{equation}
Since the junction capacitances are neglected, the charging energy 
is calculated as~\cite{Orlando99}
\begin{equation}
H_{\rm C} = \frac{1}{2C_{\rm g}} (Q_1^2 + Q_2^2) + 
V_1 Q_1 + V_2 Q_2.
\end{equation}
This energy is rewritten by using the numbers of Cooper-pairs as
\begin{equation}
H_{\rm C} = 4E_{\rm C} (n_1 - n_1^*)^2 + 4E_{\rm C} (n_2-n_2^*)^2,
\end{equation}
where $E_{\rm C}=e^2/2C_{\rm g}$, and
$n_i^* = C_{\rm g} V_i/2e$ is a normalized gate voltage.
This gives a kinetic energy for the phases. 

Here, note that
the gate voltages can control the Bloch wave-numbers. 
This can be checked by considering the wave function
$\Psi(\phi_1,\phi_2)$ satisfying the periodic condition for both
$\phi_1$ and $\phi_2$. For the alternative wave function 
$\xi(\phi_1,\phi_2)$ defined by
$\Psi(\phi_1,\phi_2) = e^{-{\rm i}n_1^* \phi_1-{\rm i}
n_2^* \phi_2} \xi(\phi_1,\phi_2)$, the kinetic part of the
Hamiltonian for $\xi(\phi_1,\phi_2)$ is modified as
\begin{equation}
\tilde{H}_{\rm C} = 4E_{\rm C} n_1^2 + 4E_{\rm C} n_2^2,
\end{equation}
while the wave function $\xi(\phi_1,\phi_2)$ satisfies the following
Bloch theorem
\begin{equation}
\xi(\phi_1+2\pi m_1,\phi_2+2\pi m_2) = e^{{\rm i}m_1 n_1^* 
+ {\rm i} m_2 n_2^*} \xi(\phi_1,\phi_2).
\end{equation}
Hence, the normalized voltages, $n_i^*$'s correspond to the 
Bloch wave-numbers for the Hamiltonian $\tilde{H} =
\tilde{H}_{\rm C} + H_{\rm J}$. We find that the feature of classical
phase dynamics for $\tilde{H}$ leads to the same mysterious normal
diffusion phenomena in SQUID's with two junctions~\cite{Tanimoto02}.
(We suppress the classical result in this Letter.)

The energy spectrum of the SQUID is obtained by diagonalizing the
Hamiltonian $H = H_{\rm C} + H_{\rm J}$ for fixed values of $n_1^*$
and $n_2^*$. We adopt the charge eigenstates as the base 
wavefunctions. Then, it is convenient to rewrite
the Hamiltonian by introducing an annihilation
operator $b_i = e^{{\rm i}\phi_i}$ satisfying an exchange
relation $[n_i,b_j]=-b_i \delta_{ij}$ as
\begin{eqnarray}
H &=& 4E_{\rm C} (n_1-n_1^*)^2 + 4E_{\rm C} (n_2 - n_2^*)^2 \nonumber
\\
&-& \frac{E_{\rm J}}{2} 
(b_1 + b^{\dagger}_1) - \frac{E_{\rm J}}{2} (b_2 + b^{\dagger}_2) 
- \frac{E_{\rm J}}{2} (b_1 b^{\dagger}_2 e^{-2\pi {\rm i} f}
+ b_2 b^{\dagger}_1 e^{2\pi {\rm i}f}).
\label{eq:Ham2}
\end{eqnarray}
The charging energy gives diagonal parts, while the junction energy
off-diagonal parts.
Here, we should note that if $f$ is an integer or half-integer,
the Hamiltonian can be represented by a real matrix. 

In this Letter, we take $E_{\rm C} = 0.002 E_{\rm J}$, and consider
the case of $f=0.5$ and $f = 0.25$. The classical phase dynamics
based on Eq.~(\ref{Ham0}) shows
chaotic behaviors in the dominant region of the 
corresponding Poincare section,
when the energy is taken between the saddle-point potential 
energy $E_{\rm sad}$ and the potential-top energy $E_{\rm top}$.
Here, the spectral properties are studied in the range 
$E_{\rm sad} \le E \le E_{\rm sad} + 0.75(E_{\rm top}-E_{\rm sad})$.
The ordinary unfolding procedure is performed to obtain 
the level statistics~\cite{Haake00,Mehta91}.
Depending on the Bloch wave-numbers, $n_1^*$ and $n_2^*$, the energy spectra
show various statistics: the Poisson-like, GOE, and GUE statistics. 
To see this wave-number dependence, we fit the spectra to the
Izrailev distribution~\cite{Izrailev88}
\begin{equation}
P(S) = A \left(\frac{\pi S}{2}\right)^{\beta} \exp \left[
-\frac{1}{16} \beta \pi^2 S^2 - \left( B-\frac14 \pi \beta \right) S 
\right],
\end{equation}
which interpolates among different universality classes by the
parameter $\beta$. The parameters, $A$ and $B$ are normalization constants.
One can recover the Poisson distribution for $\beta = 0$, while
it agrees with the GOE (GUE) distribution for $\beta =1$ 
($\beta = 2$) on the 5 percent level. Although the Izrailev distribution is 
originally proposed for localization problems, we here use it for
qualitative characterization of the spectra.

\begin{figure}[tbh]
\begin{center}
\includegraphics[width=65mm]{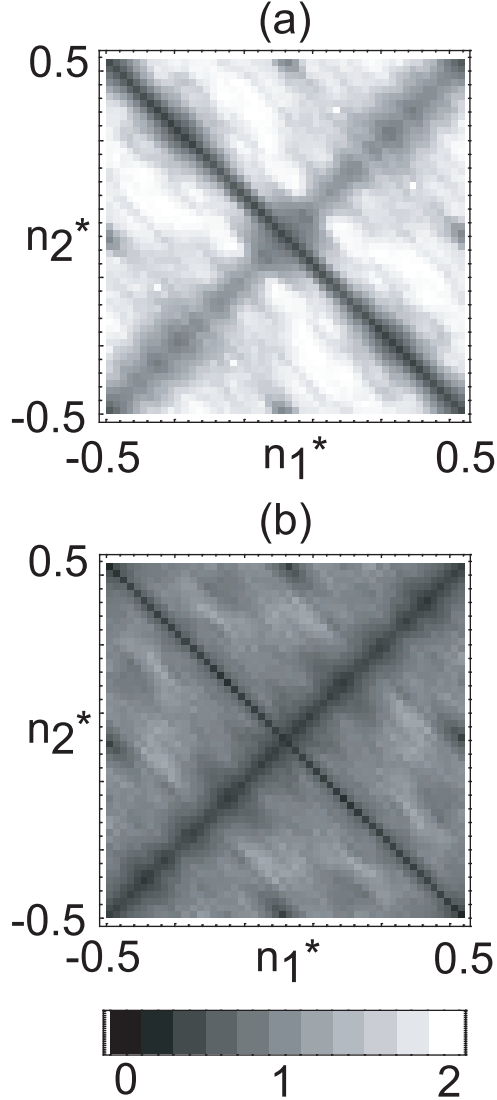}
\end{center}
\caption{The density plot of the parameter $\beta$ obtained
by fitting the spectrum to the Izrailev distribution: (a) $f = 0.25$ and (b)
$f = 0.5$. $\beta=0$ (the darkest) and $\beta = 2$ (the brightest).}
\label{fig:diagram}
\end{figure}

\begin{figure}[tbh]
\begin{center}
\includegraphics[width=65mm]{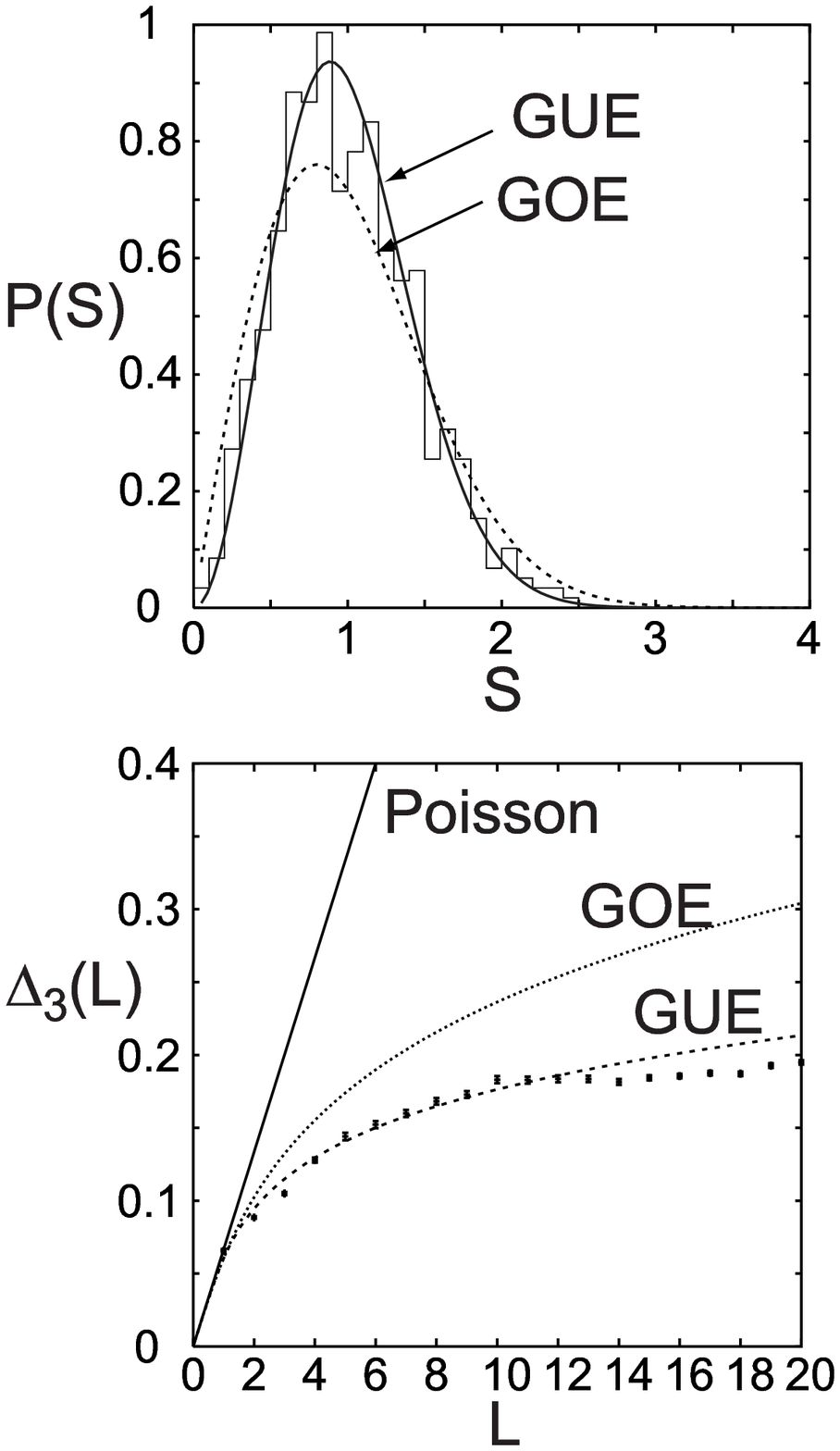}
\end{center}
\caption{(a) The nearest-neighbor level distribution $P(S)$ and
(b) the spectral rigidity $\Delta_3(L)$ for $f = 0.25$. The normalized
gate voltages are taken as $(n_1^*,n_2^*) = (0.25,0)$.}
\label{fig:GUE}
\end{figure}

\begin{figure}[tbh]
\begin{center}
\includegraphics[width=65mm]{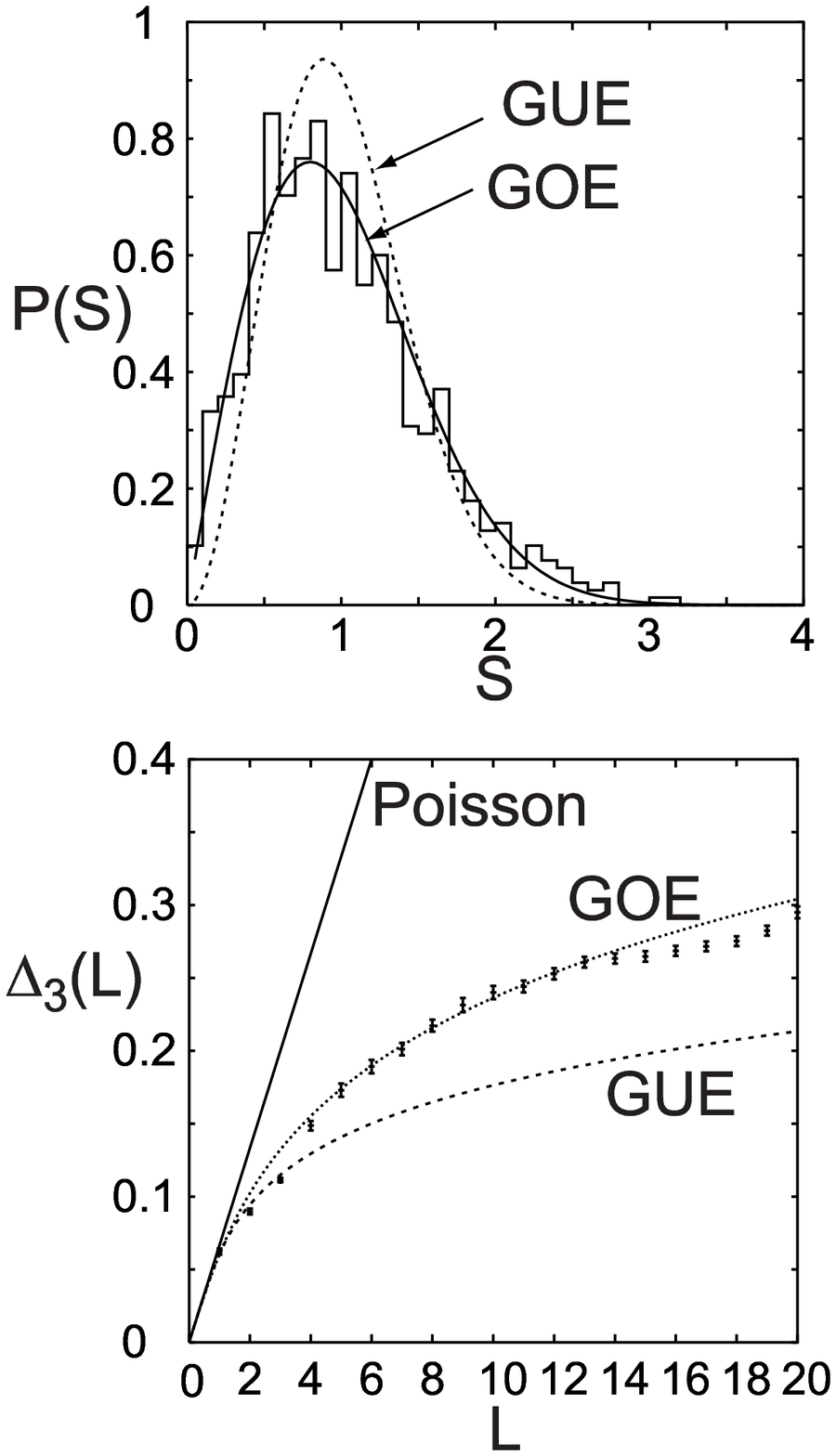}
\end{center}
\caption{(a) The nearest-neighbor level distribution $P(S)$ and
(b) the spectral rigidity $\Delta_3(L)$ for $f = 0.5$. The normalized
gate voltages are taken as $(n_1^*,n_2^*) = (0.25,0)$.}
\label{fig:GOE}
\end{figure}

The density plots of the fitted values of $\beta$ in the 
first Brillouin zone (FBZ) of the $n_1^*$-$n_2^*$ plane are shown
in Fig.~\ref{fig:diagram}. In the case of $f = 0.25$, the spectra
shows the GUE distribution ($\beta = 2$)
except for the symmetry lines, $n_1^* = n_2^*$
and $n_1^* = - n_2^*$, and the points, $(n_1^*,n_2^*)=(\pm 0.5,0)$ and
$(0,\pm 0.5)$. In the case of $f = 0.5$, the spectra shows the GOE
distribution ($\beta = 1$) except the same symmetry lines and points. 
Let us first discuss the universality class of the region, which
is on neither the symmetry lines nor points. In the case of $f = 0.25$,
the Hamiltonian includes the imaginary number in the last term
in (\ref{eq:Ham2}). Hence, it is expected that 
the time-reversal symmetry is broken, and the level statistics
is governed by the GUE. Actually, the nearest
neighbor distribution $P(S)$ and spectral rigidity $\Delta_3(L)$
agree well with the GUE statistics as shown in Fig.~\ref{fig:GUE}.
In the case $f = 0.5$, the Hamiltonian can be expressed by a
real symmetric matrix as seen in (\ref{eq:Ham2}). 
Hence, the system exhibits the time-reversal symmetry, and
the level statistics belongs to the GOE. Actually, the result for
$P(S)$ and $\Delta_3(L)$ agrees with the GOE statistics as shown
in Fig.~\ref{fig:GOE}. Since the value of $f$ can be varied
by the external magnetic flux, the universality class can be
easily controlled in this SQUID system.

Thus, the signature of quantum chaos has been obtained in the SQUID
system. So, the main purpose of this Letter has been achieved.
It, however, remains to explain the deviations from the expected 
universal class on the symmetry lines and points 
as seen in Fig.~\ref{fig:diagram}. We end this Letter by 
brief discussion on this secondary feature in the SQUID system.

First, we consider the case of $f=0.25$. We divide the regions where 
the deviation from the GUE statistics is seen into five parts: 
(A) the line $n_1^*=n_2^*$ ($n_1^* \neq 0$),
(B) the line $n_1^* = -n_2^*$ ($n_1^* \neq 0$), 
(C) the point $(n_1^*,n_2^*)= (0,0)$,
(D) the points $(n_1^*,n_2^*) = (\pm 0.5,0)$, and
(E) the points $(n_1^*,n_2^*) = (0,\pm 0.5)$.
The deviation from the GUE statistics in each region 
can be explained by corresponding additional symmetries
characterized by operators commuting with the Hamiltonian.
In the region (A), it is crucial to consider 
the operator $P_1$ defined by $| n_2, n_1 \rangle = 
P_1|n_1, n_2 \rangle$, where
$| n_1, n_2 \rangle$ is the eigenkets of the Cooper-pair-number 
operator. It can be checked that the operator $P_1 T$ 
commutes with the Hamiltonian on the line $n_1^*=n_2^*$, where
$T$ is the time-reversal operator. Then, the spectral
statistics is governed by the GOE statistics by the discussion
of the false time-reversal symmetry~\cite{Robnik86}.
Similarly, in the region (B), the mirror-reversal operator
$P_2$ defined by $|-n_2,-n_1\rangle =P_2 | n_1, n_2 \rangle$
commutes with the Hamiltonian. Then, the eigenkets can be classified
into two kinds by the parity defined as the eigenvalue
of the operator $P_2$. The spectrum taken only for even(odd) states
is expected to show the level repulsion. The level correlation between
odd and even states is, however, suppressed because the matrix 
elements of the Hamiltonian $\langle \psi_i | H | \psi_j \rangle$ is zero
when $\psi_i$ and $\psi_j$ have different parity. As a result,
the spectrum become the mixture of two independent GUE statistics.
This mixed level statistics is denoted in this Letter with 
$2$-GUE~\cite{Xu98,Xu01}. In the region (C), both $P_1 T$ and $P_2$ 
commutes with the Hamiltonian. Then, the spectrum becomes the mixture
of the two independent GOE statistics ($2$-GOE). In the region (D),
a kind of the mirror reversal symmetries exists. The symmetry operator
$P_3$ is defined as $| 1-n_1, -n_2\rangle = P_3 |n_1,n_2 \rangle$, 
and $P_3 T$ commutes with the Hamiltonian. Then, the spectral statistics 
obeys the GOE by the same reason as the region (A).
Similarly, in the region (E), $P_4 T$ commutes with the Hamiltonian,
where $P_4$ is defined as $| -n_1, 1-n_2 \rangle = P_4 | n_1,n_2 \rangle$.
Hence, the level statistics is described by the GOE.

Next, we consider the case of $f=0.5$. The deviation from the GOE
statistics is observed near the same lines and points as the case of
$f=0.25$.
The explanation of this deviation is quite parallel to the previous
paragraph. In the region (A) and (B), the operator $P_1$
and $P_2$ commute with the Hamiltonian, respectively.
Hence in these regions, the level spectrum becomes the mixture
of two independent GOE statistics ($2$-GOE). In the region (C),
because both $P_1$ and $P_2$ commute with the Hamiltonian,
the spectrum obeys the $4$-GOE statistics. In the region (D) and (E),
the spectrum is described by the $2$-GOE statistics, 
since $P_3$ and $P_4$ commute with the Hamiltonian, respectively.

In summary, spectral statistics in band structures 
has been discussed in a realistic SQUID model. 
This system can change the Bloch wave-numbers by controlling
the gate voltages. The normalized external flux $f$
can control the time-reversal symmetry of the system, and
the crossover from the GUE to the GOE can be observed 
by varying $f$.
By taking the circuit parameters as symmetric, the effect of
the secondary discrete symmetries can also be discussed. On specific
lines and points in the first Brillouin zone, the spectra may
obey different statistics determined by the operator characterizing
the discrete symmetry of the system. We expect that this work can 
be a starting
point to consider the quantum chaos in a well-controlled system.
A variety of dynamical features of quantum chaos can also be observed
in the quantum dynamics by using the excellent ability of quantum 
manipulation and measurement in the SQUID system, which will be
investigated in due course.

The authors thank A. Terai for useful discussion and 
critical reading of the manuscript.

\end{document}